\documentclass[a4paper,11pt]{article}
\usepackage{pos}

\title{2p-2h Cross-Section Systematics in DUNE} 
\subtitle{Contribution to the 25th International Workshop on Neutrinos from Accelerators}

\author[a,*]{Lars Bathe-Peters}
\collab{for the DUNE collaboration}
\affiliation[a]{University of Oxford, Oxford, OX1 3RH, United Kingdom}

\emailAdd{lars.bathe-peters@ox.ac.uk}

\abstract{For the operation of precision neutrino experiments, the understanding of neutrino interactions with matter is a preconditioned requirement for all detections and measurements of neutrinos. The largest uncertainties in estimating neutrino-nucleus interaction cross sections arise from the incomplete understanding of nuclear effects. In the study of neutrino oscillations and nuclear scattering processes, obtaining an interaction model with associated uncertainties is of sub- stantial interest for the neutrino physics community. This report presents studies of simulated CC 2p-2h interactions, in which a neutrino interacts with a bound pair of nucleons. This interaction mode is very poorly constrained by current data. A comparison of three leading CC 2p-2h models is presented, along with a number of uncertainty parameters that have been implemented to account for model-to-model discrepancies in the DUNE oscillation analysis.}

\FullConference{NuFact 2024 - The 25th International Workshop on Neutrinos from Accelerators \\
 16-21, September 2024\\
Argonne National Laboratory, Lemont, Illinois, United States\\}

\begin{document}
\maketitle

\section{Introduction}\vspace{-4mm}
    \begin{figure}[hbt]
	\centering
	\includegraphics[width=\textwidth]
    {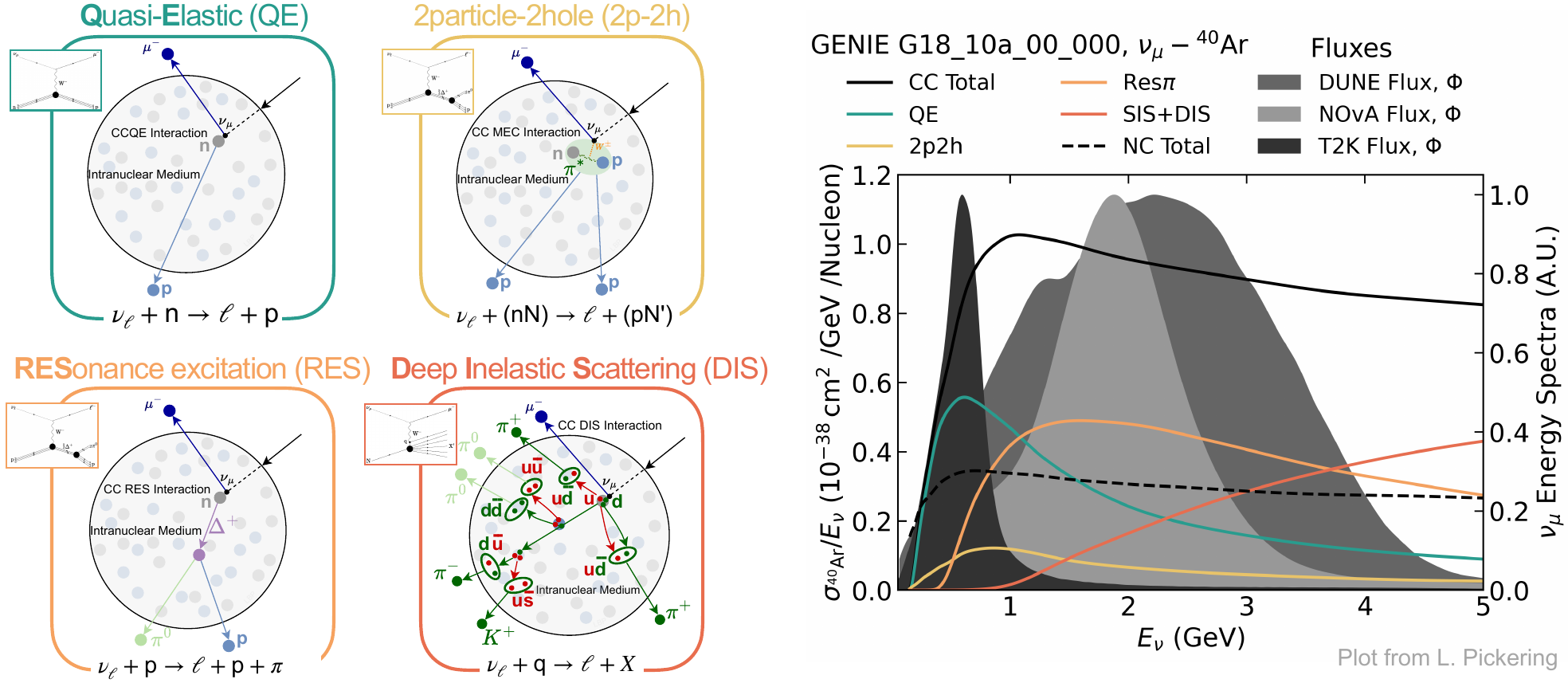}
	\caption[nuintscale]{Left: Illustrations of neutrino-nucleus interactions. Figures taken or adapted from \cite{bathe-peters2020} Right: Unoscillated DUNE, No$\nu$a and T2K flux predictions overlaid with neutrino-argon cross sections as a function of the neutrino energy. Plot from L. Pickering.}
	\label{nu_int_scale}
    \end{figure}
    \begin{figure}[bth]
	\centering
	\includegraphics[width=.9\textwidth]{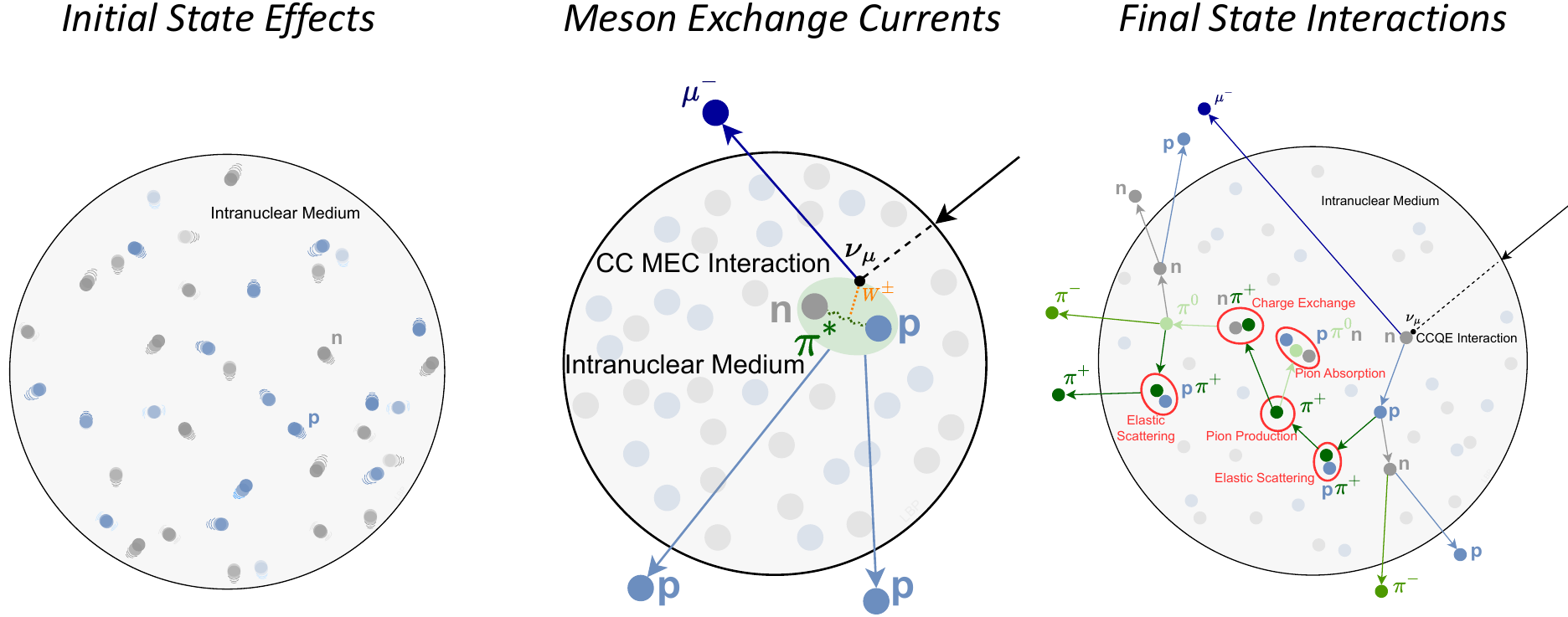}
	\caption[nucleareffects]{Nuclear Effects that complicate the description of neutrino-nucleus scattering. These can be categorised into initial state effects (left) such as Fermi motion, nucleon-binding processes (middle) for example via virtual pion exchange and final-state interactions (right) such as pion production or charge exchange within the nuclear medium. Figures taken or adapted from \cite{bathe-peters2020, bathe-peters2022}.}
	\label{nuclear_effects}
    \end{figure}
    \begin{figure}[hbt]
	\centering
	\includegraphics[width=.7\textwidth]{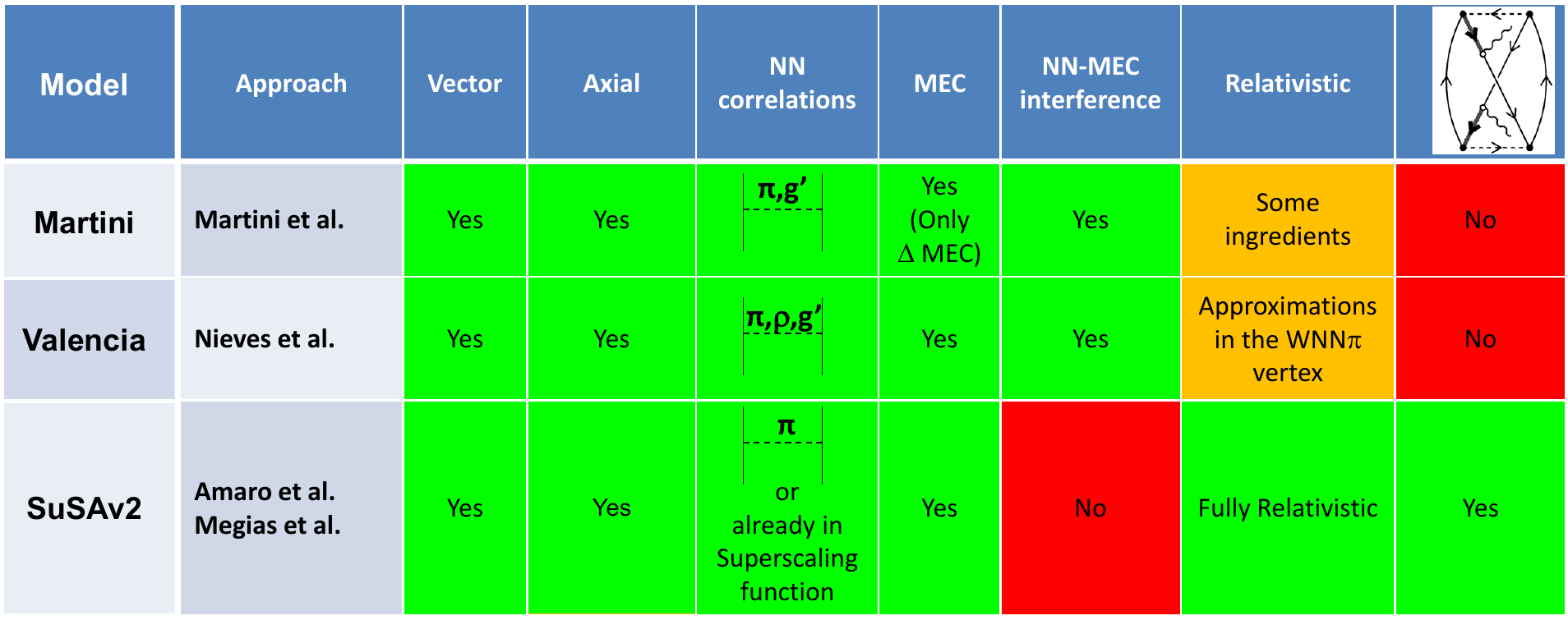}
	\caption[2p2hmodels]{Charged-current 2p-2h model differences in the Martini, Valencia and the SuSAv2 models. Table adapted from \cite{martini2016}.}
	\label{2p2h_models}
    \end{figure}
    \vspace{-4mm}
The DUNE experiment is a $1300$km long-baseline neutrino oscillation experiment with the goal of measuring neutrino oscillation parameters, searching for CP violation, proton decay and supernova neutrinos. Moreover, there is a physics program for beyond the standard model physics, atmospheric neutrinos, dark matter searches as well as a rich neutrino interaction program. One of the roles of the DUNE near detector is the tuning of the neutrino-nucleus interaction model. A determination of neutrino oscillation parameters requires an accurate reconstructed neutrino energy, since the relevant cross section applies to the true neutrino energy. The right plot in \cref{nu_int_scale} shows the unoscillated DUNE near detector flux predictions alongside the oscillated one overlaid with the partial neutrino cross section predictions for various neutrino interaction channels shown on the left. This picture becomes more complicated when considering nuclear effects (see \cref{nuclear_effects}). 
\vspace{-5mm}
\begin{table}[b]
\centering
\caption{Model elements of tunes used in the simulation of muon neutrino interactions on argon.} \label{tab:new_tunes}
\resizebox{.8\textwidth}{!}{
  \begin{tabular}{|l||c|c|c|}
        \hline
		\diagbox{\textbf{Model element}}{\textbf{Set name}} & $\text{G18\_10a\_00\_000}$ & $\text{G18\_10s\_00\_000}$ &  $\text{G18\_10e\_00\_000}$ \\ \hline \hline
		Nuclear (Ground-State) Model	& Local Fermi Gas & Local Fermi Gas & Local Fermi Gas \\ \hline
        Quasi-Elastic (QE) processes &  Nieves & Nieves & Nieves \\ \hline
		\cellcolor{yellow!20}2p-2h (MEC)-processes	& \cellcolor{yellow!20}Nieves (Valencia) & \cellcolor{yellow!20}SuSAv2 & \cellcolor{yellow!20}Empirical \\ \hline
		Resonance (RES) production	& Berger-Sehgal & Berger-Sehgal & Berger-Sehgal \\ \hline
  		Deep Inelastic Scattering (DIS)	& Bodek-Yang & Bodek-Yang & Bodek-Yang \\ \hline
		Coherent (COH) production	& Berger-Sehgal & Berger-Sehgal & Berger-Sehgal \\ \hline
		Final-State Interactions (FSI)	& INTRANUKE hA 2018 & INTRANUKE hA 2018 & INTRANUKE hA 2018 \\ \hline
  \end{tabular}
}
\end{table}
\section{Charged-Current 2p-2h Neutrino-Nucleus Interactions}\vspace{-2mm}
2p-2h Charged-Current (CC) neutrino-nucleus scattering events are interactions in which the neutrino scatters off a bound state of two nucleons giving rise to a lepton and two particles in the final state. As two particles get ejected from the nuclear medium, they leave two holes behind, hence the name 2-particle-2-hole interactions. 
The differences in theoretical computations of the 2p-2h process according to the 2p-2h Martini, Valencia and SuSAv2 models are highlighted in \cref{2p2h_models}. The main differences between these models lie in the contributing Feynman diagrams that enter the calculation and the interferences of respective resonances. 
\vspace{-1mm}
\section{CC 2p-2h Uncertainty Parameters}\vspace{-2mm}
There are currently three 2p-2h models implemented in the GENIE neutrino event generator. The model set configurations used for all results shown in this work are listed in \cref{tab:new_tunes}.
The following results were obtained by simulating muon-neutrino interactions on argon using the GENIE neutrino event generator for the CC 2p-2h SuSAv2, Valencia and Empirical model. The goal of this study is to choose uncertainties such that the measurement of the oscillation parameters is not biased in case the wrong model is chosen. Hereby, systematic parameters allow us to challenge and depart from the physics assumptions implemented in the central value predictions which in DUNE is the CC 2p-2h SuSAv2 model. This is achieved by the method of event reweighting.
\vspace{-2mm}
\subsection{Normalisation Parameter}
\begin{figure}[htb!]
	\begin{subfigure}[b]{.27\textwidth}
        \includegraphics[width=\textwidth]{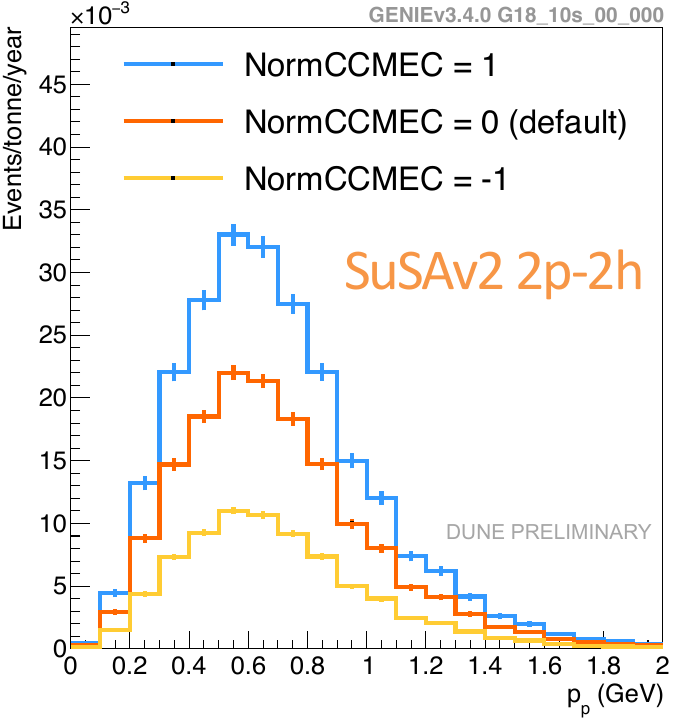}
    \end{subfigure}
    \hfill
	\begin{subfigure}[b]{.28\textwidth}
        \includegraphics[width=\textwidth]{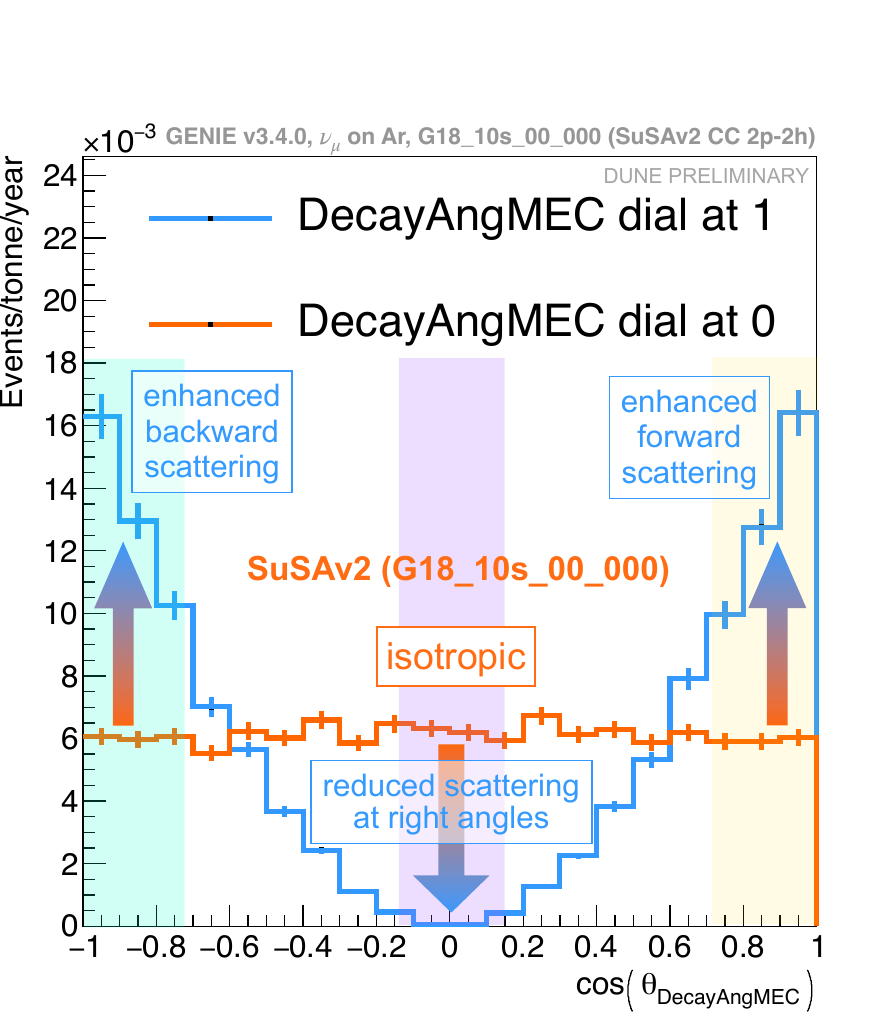}
    \end{subfigure}
    \hfill
	\begin{subfigure}[b]{.27\textwidth}
        \includegraphics[width=\textwidth]{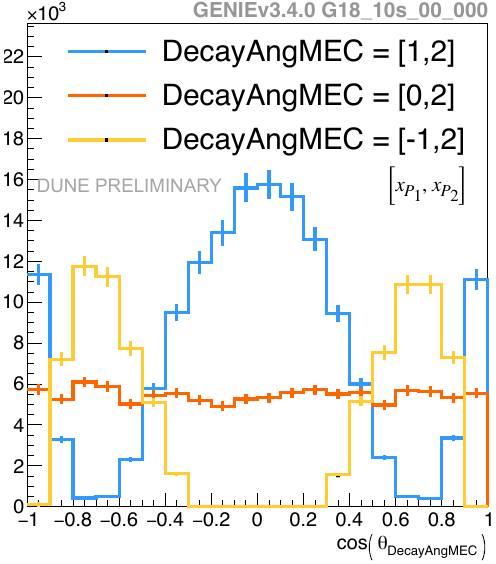}
    \end{subfigure}
 	\caption[normangleparams]{Left: The normalisation parameter. The default CC2p-2h SuSAv2 model prediction is shown in orange. Middle and right: Reweighting by changing the nucleon angular distribution dependency.}
	\label{norm_angle_params}
    \end{figure}

\vspace{-2mm}
As the name suggests the normalisation parameter changes the absolute normalization of variable distributions. The effect of varying its related dial can be seen on the left in \cref{norm_angle_params}. 
 By changing the normalisation parameter, one can reweight these distributions up and down.
\vspace{-2mm}
\subsection{Nucleon Angular Distribution}\vspace{-2mm}
A slightly more complicated parameter is a nucleon angular distribution parameter that modifies the angular dependency of the struck nucleon pair. 
In the default model prediction (SuSAv2), the nuclear angular distribution is isotropic, meaning that the cosine of this angle is a flat distribution. This can be seen in in the middle and right in \cref{norm_angle_params}. This assumption is challenged by introducing a weight that leads to an angular distribution that is dependent on the squared-cosine of the decay angle. The effect is an enhancement of events in the forward and backward scattering regions and a suppression of scattering at right angles. This choice of changing the angular dependence in this way is not well physically motivated as the underlying physics of these outgoing nucleon directions is challenging to describe. In order to accommodate this difficulty, another dial was implemented that changes the frequency of the distribution (see \cref{norm_angle_params}). 
\vspace{-2mm}
\subsection{Nucleon Pair Content}
\begin{figure}[htb!]
	\begin{subfigure}[b]{.32\textwidth}
        \includegraphics[width=\textwidth]{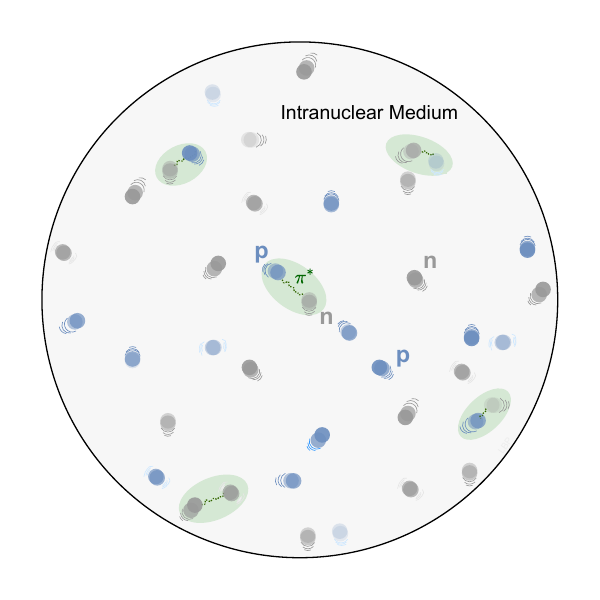}
    \end{subfigure}
    \hfill
	\begin{subfigure}[b]{.25\textwidth}
        \includegraphics[width=\textwidth]{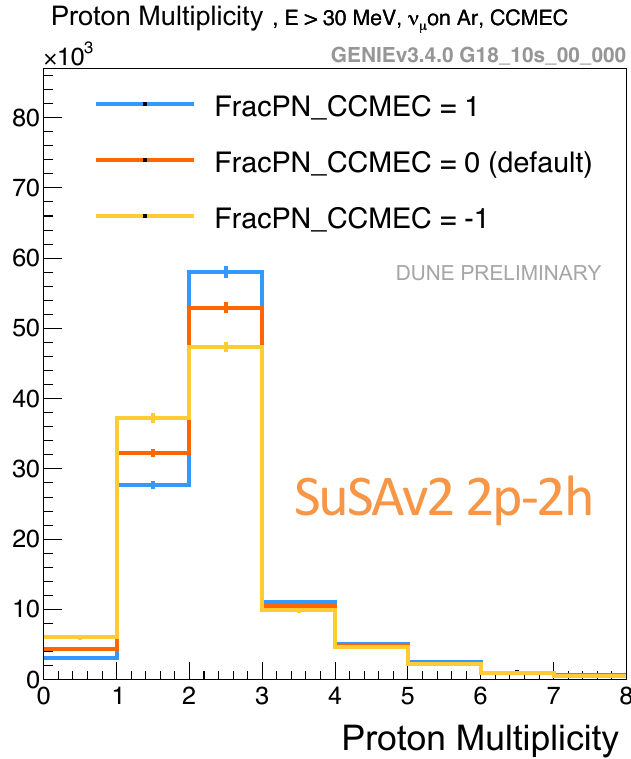}
    \end{subfigure}
    \hfill
	\begin{subfigure}[b]{.25\textwidth}
        \includegraphics[width=\textwidth]
        {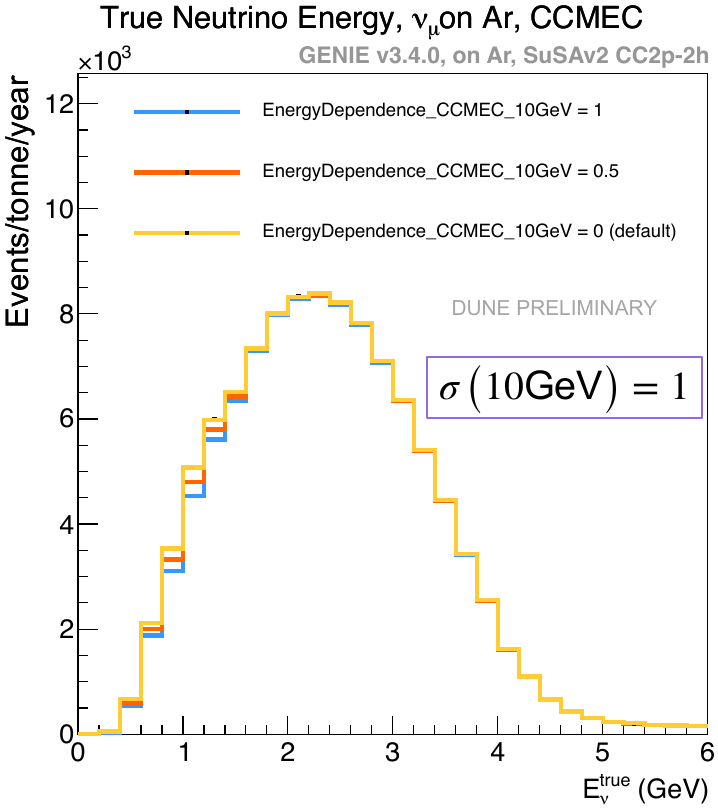}
    \end{subfigure}
 	\caption[pnfrac]{Left: A default fraction of pn-pairs inside the nuclear medium prior to the neutrino interaction determines the frequency of neutrinos scattering off bound states of two nucleons. These give rise to different proton multiplicites (middle). Right: Effect of reweighting the energy dependence parameter.}
	\label{pnfrac}
    \end{figure}

\vspace{-2mm}
The nucleon pair content parameter changes the fraction of pn-pairs inside the nuclear medium (see \cref{pnfrac}).
As the number of np-pairs is increased, there are more 2-proton final states. Consequently, there are less 1-proton final states as more np-pairs means less nn-pairs. 
\vspace{-2mm}
\subsection{CC 2p-2h Model Shape Differences}
\begin{figure}[htb!]
	\begin{subfigure}[b]{.31\textwidth}
        \includegraphics[width=\textwidth]{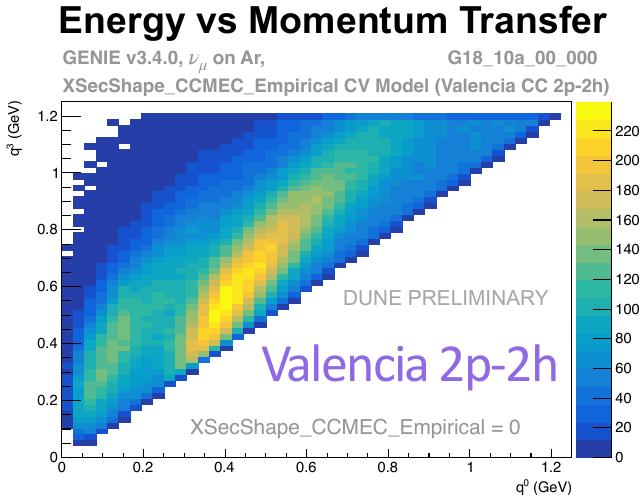}
    \end{subfigure}
    \hfill
	\begin{subfigure}[b]{.33\textwidth}
        \includegraphics[width=\textwidth]{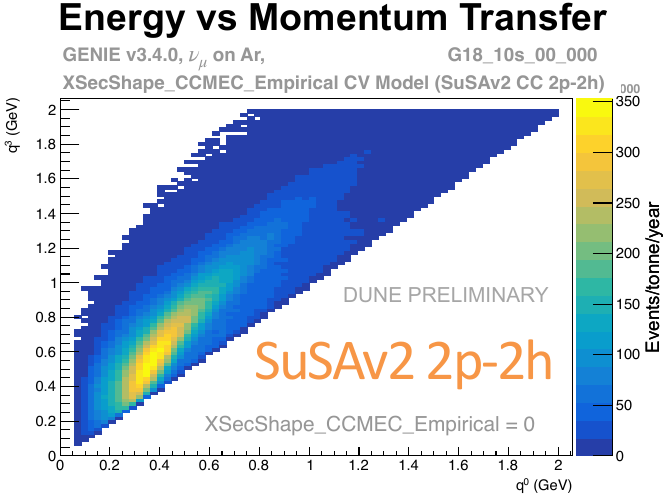}
    \end{subfigure}
    \hfill
	\begin{subfigure}[b]{.33\textwidth}
        \includegraphics[width=\textwidth]
        {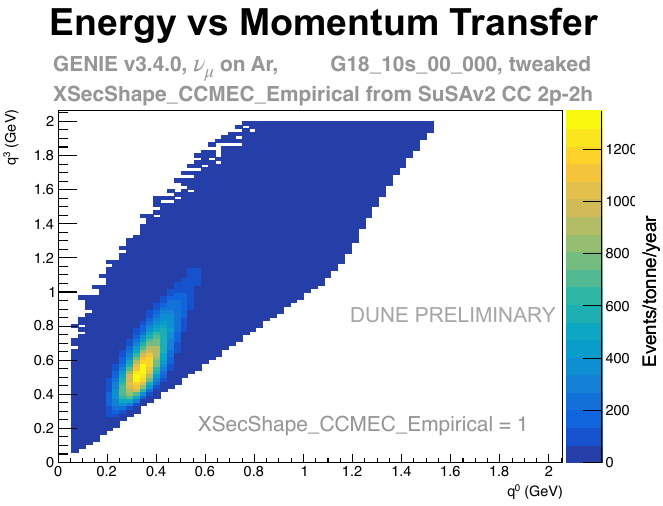}
    \end{subfigure}
	\caption[xsecshapeemp-sus]{The cross-section interpolation parameter allows the transition from the Valencia (left) and SuSAv2 (middle) model to the Empirial (right) model via event reweighting.}
	\label{xsecshape_emp-sus}
    \end{figure}

\vspace{-2mm}
The cross-section shape parameter interpolates between the CC 2p-2h models. This is usually considered in the three-momentum transfer versus energy-transfer plane as it is shown in \cref{xsecshape_emp-sus}.
%
The idea is to reweight from one to the other model. The reweighting using this parameter 
from the SuSAv2 to the Empirical model is shown in \cref{xsecshape_emp-sus}. One characteristic feature of the Valencia model in this plane is the second peak that results from resonances and interference terms. This two-peak structure vanishes when reweighting from the Valencia to the empirical model. There is good agreement with the nominal prediction. In the future it would be interesting to weight features of the models individually rather than interpolate between models.
\vspace{-2mm}
\subsection{Energy Dependence}
\begin{figure}[htb!]
    \centering
	\begin{subfigure}[b]{.32\textwidth}
        \includegraphics[width=\textwidth]{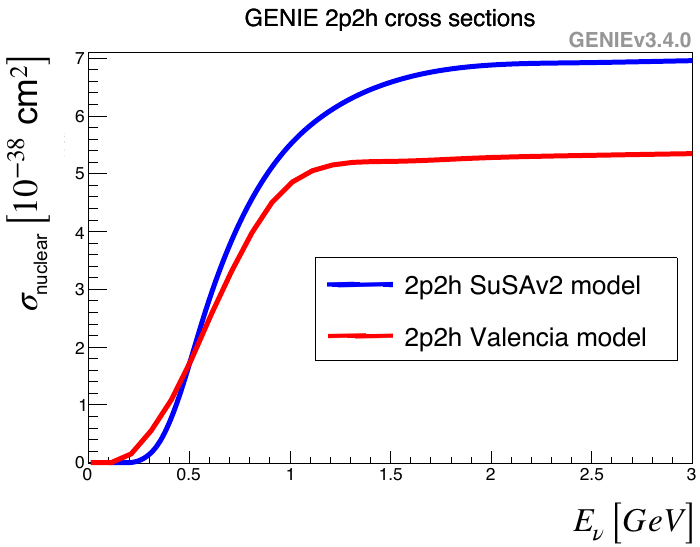}
    \end{subfigure}
	\begin{subfigure}[b]{.32\textwidth}
        \includegraphics[width=\textwidth]{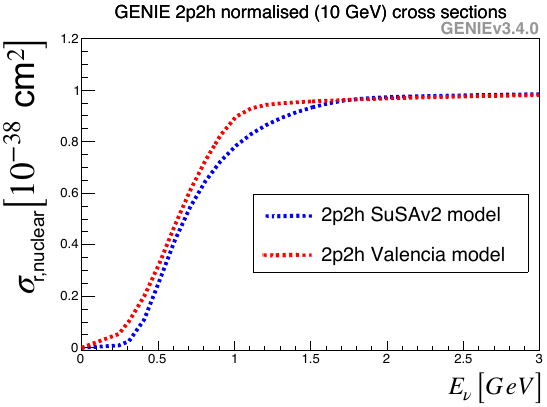}
    \end{subfigure}
 	\caption[energydep]{Left: The different CC 2p-2h model predictions by the Valencia and the SuSAv2 models as a function of the neutrino energy. Right: The renormalised cross-section predictions for an anchoring point of $10$GeV.}
	\label{energy_dep}
    \end{figure}

\vspace{-2mm}
The energy dependence parameter changes the energy dependence of the 2p-2h cross sections. Different models have different predictions on the energy dependence of the cross section (see left in \cref{energy_dep}).
In order to renormalise the cross section an anchoring point can be chosen to make the different cross-section predictions coincide at certain energies. In order to put an emphasis on the shape differences, the anchoring point was chosen at a relatively high energy, at $10$ GeV. The renormalised cross section can be seen in the middle of \cref{energy_dep}. By computing the ratio between the higher versus the lower cross-section value, one can get a handle on how to reweight the neutrino energy distribution. The reweighted neutrino-energy distribution is shown on the right in \cref{pnfrac}. The difference between the nominal and the reweighted distribution is most prominent for energies smaller than $1.5$ GeV. 
\vspace{-2mm}
\section{Summary and Outlook}\vspace{-2mm}
A summary of the charged-current 2p-2h uncertainty parameters that will be available in GENIEReweight are shown in \cref{summary_table}. There is a plethora of novel ideas regarding new dials or variations in the existing ones. Concerning the nuclear decay angle parameter, there is currently no implementation that considers the hadron kinematics dependence on the momentum transfer. Another idea is to consider a parameter that addresses the uncertainties on the removal energy carried away by the struck nucleon pair. Furthermore, the reweighting tools should be enhanced to enable reweighting for anti-neutrino scattering, especially for the cross-section interpolation parameter and the energy- dependence parameter. Regarding the the energy- dependence parameter, a more sophisticated approach is to reweight the structure functions directly in the implementation of the cross section. In that regard, a better agreement between the theoretical predictions and the implementations in the Monte Carlo event generators is essential \cite{bib:th_gen}. 

As neutrino-nucleus interactions remain very complex and necessitate dedicated study, there is substantial need to develop fit parameters to estimate systematic uncertainties in order to choose uncertainties in a way such that the measurements of the neutrino oscillation parameters are not biased. This will allow a robust estimate of systematic uncertainties in modern and future neutrino oscillation experiments such as DUNE.
    \begin{table}[hbt]
   	\caption[summarytable]{For central values and uncertainties in MicroBooNE, see \cite{bib:uboone_dials}. Table adopted from Table VIII in \cite{bib:uboone_dials}.}
	\label{summary_table}
	\centering
	\includegraphics[width=\textwidth]{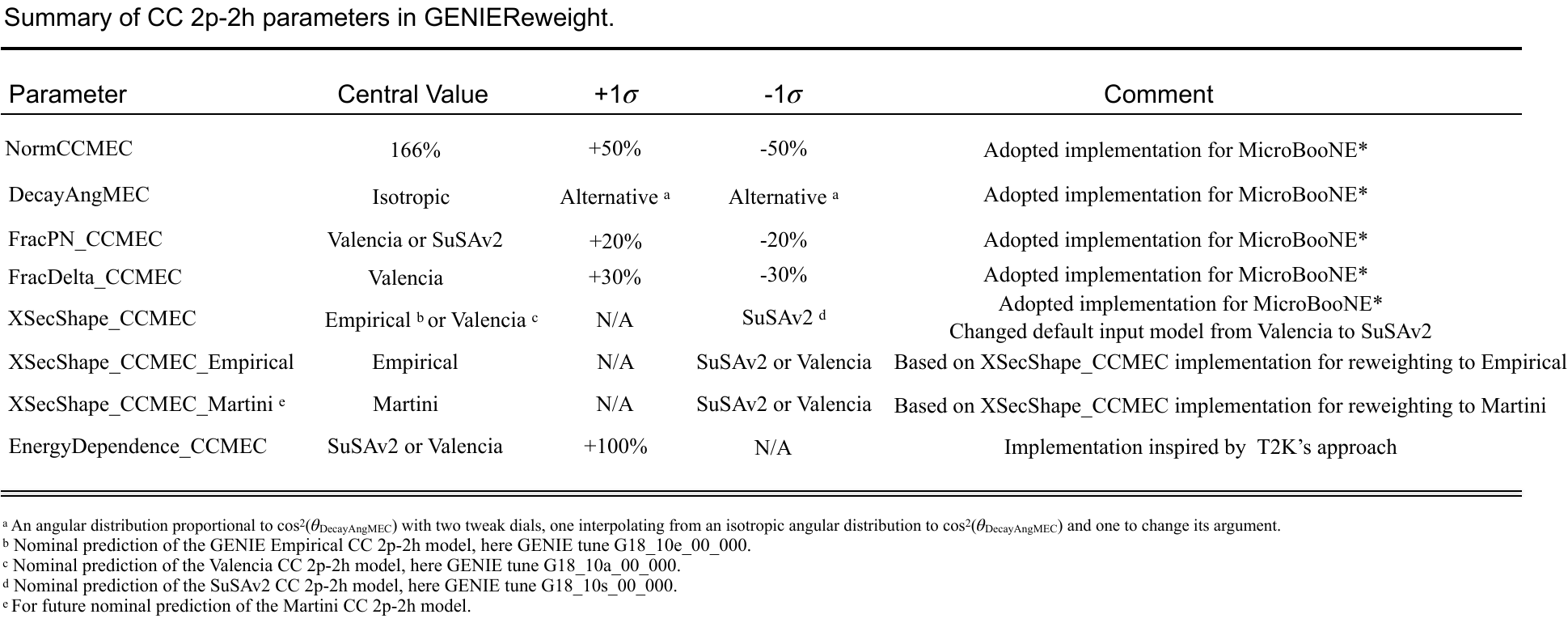}
    \end{table}


\vspace{-3mm}
\begin{thebibliography}{99}
\begin{small}
\setlength{\itemsep}{0pt}   
\setlength{\parsep}{0pt}    

\bibitem{bathe-peters2020}
L. Bathe-Peters, \emph{Studies of Single-Transverse Kinematic Variables for Neutrino Interactions on Argon}, \href{https://www.osti.gov/biblio/1661674}{\emph{FERMILAB-MASTERS-2020-03}}.

\bibitem{bathe-peters2022}
L. Bathe-Peters \textit{et al}, \emph{Comparing generator predictions of transverse kinematic imbalance in neutrino-argon scattering}, \href{https://arxiv.org/pdf/2201.04664}{\emph{arXiv:2201.04664}} [hep-ph].

\bibitem{martini2016}
M. Martini \textit{et al}, \emph{Inclusion of multi-nucleon effects in RPA-based calculations for -nucleus scattering}, \href{https://esnt.cea.fr/Phocea/Page/index.php?id=59}{\emph{ ESNT 2p-2h workshop}} (2016).

\bibitem{bib:th_gen}
J. E. Sobczyk \textit{et al}, \emph{Exclusive-final-state hadron observables from neutrino-nucleus multinucleon knockout}, \href{https://journals.aps.org/prc/abstract/10.1103/PhysRevC.102.024601}{\emph{Phys. Rev. C 102, 024601 (2020)}} [nucl-th].

\bibitem{bib:uboone_dials}
P. Abratenko \textit{et al}, \emph{New CC$0\pi$ GENIE model tune for MicroBooNE}, \href{https://journals.aps.org/prd/abstract/10.1103/PhysRevD.105.072001}{\emph{Phys. Rev. D 105, 072001 (2022)}} [hep-ex].

\end{small}
\end{thebibliography}
\end{document}